\documentclass[prc,aps,showpacs,nofootinbib,twocolumn]{revtex4}
\usepackage{epsfig}
\usepackage{amssymb}
\usepackage{amsmath}
\usepackage{amsfonts}
\begin{document}
\title{Shell-model Hamiltonians from Density Functional Theory}
\author{Y. Alhassid$^{1}$, G.F. Bertsch$^{2}$, L. Fang$^{1}$ and B. Sabbey$^{2}$
}
\affiliation{$^{1}$Center for Theoretical Physics, Sloane Physics
Laboratory, Yale University, New Haven, CT 06520\\
$^{2}$Department of Physics and Institute of Nuclear Theory,
Box 351560\\
University of Washington
Seattle, WA 98915\\}
\def\rhob{{\boldsymbol\rho}}
\def\hb2{{\bf h}^{(0)}}
\def\PrhoP{{\bf P \rhob_{\rm mf} P}}
\def\be{\begin{equation}}
\def\ee{\end{equation}}
\def\Fe{$^{56}$Fe}
\def\Cr{$^{66}$Cr}
\def\Dy{$^{162}$Dy}
\def\Ne{$^{20}$Ne}
\def\Mg{$^{24}$Mg}
\def\Si{$^{28}$Si}
\def\S{$^{32}$S}
\def\Ar{$^{36}$Ar}
\def\lb{\langle}
\def\rb{\rangle}
\def\ajm{\alpha jm}
\def\Tr{{\rm Tr}\,}
\def\tr{{\rm tr}\,}
\newcommand{\nuc}[2]{$^{#1}${#2}}
\begin{abstract}
The density functional theory of nuclear structure provides a
many-particle wave function that is useful for static
properties, but an extension of the theory is necessary to
describe correlation effects or other dynamic properties. Here
we propose a procedure to extend the theory by mapping the
properties of the self-consistent mean-field Hamiltonian onto an
effective shell-model Hamiltonian with two-body interactions. 
In this initial study, we consider the $sd$-shell nuclei \Ne,
\Mg, \Si, and \Ar. Our first application is in the framework of
the USD shell-model Hamiltonian, using its mean-field
approximation to construct an effective Hamiltonian and
partially recover correlation effects.  We find that more than
half of the correlation energy is due to the quadrupole
interaction.  We then follow a similar procedure but using the
SLy4 Skyrme energy functional as our starting point and
truncating the space to the spherical $sd$ shell.  The
constructed shell-model Hamiltonian is found to satisfy minimal
consistency requirements to reproduce the properties of the
mean-field solution.  The quadrupolar correlation
energies computed with the mapped Hamiltonian are reasonable compared 
with those computed by other methods.
\end{abstract}

\maketitle

\section{Introduction}

Quantitative theory of nuclear structure has followed two different
paths. One path is a self-consistent mean-field theory (SCMF)~\cite{heenen},
also called density functional theory,\footnote{This terminology is used because of the similarity to the density functional theory of many-electron systems~\cite{kohn}.} and the other is
the configuration-interaction shell model (CISM) approach~\cite{brown88}. The former
requires only a few parameters and has broad applicability across the
table of nuclides, while the latter is specific to regions of
the nuclides and as usually formulated requires many parameters
to specify the Hamiltonian.  However, the SCMF is of limited
accuracy for calculating binding energies, and the treatment 
of excitations requires extensions of the theory. 
 In contrast, the CISM gives a rather accurate and
almost\footnote{The qualification is due to the common occurrence of
intruder states.} complete description of the low-energy spectroscopy.
Brown and Richter~\cite{br98} proposed to use the SCMF to determine
the single-particle Hamiltonian of the CISM so as to gain advantages of both
approaches.  In this work we go farther, using 
the SCMF to also determine parts of the two-body interaction Hamiltonian.

Our work has several motivations.  One is to develop a
well-justified and accurate theory of nuclear masses.
SCMF is systematic with a strong theoretical foundation, but it may 
have reached
its limits of accuracy at a level of an MeV rms residual error
over the table of known masses.  Some effects of correlations are
missing from the theory,  and indeed better accuracy can be achieved
when phenomenological terms are included to take them into 
account~\cite{goriely}.  
A systematic calculation of those correlation energies would be
possible if we could find a way to map the SCMF on a shell-model 
Hamiltonian\footnote{The SCMF energy functionals in 
use are fitted to the binding energies, and in principle they
should be refitted if the correlation energies is added as
a separate contribution.  In practice, the refit does not change
the parameters significantly so the original functional may be
used in the determination of the Fock-space Hamiltonian.}.

Another motivation is to bring more precision to shell-model
calculations of one-body observables such as quadrupole moments and transition
intensities.  For example, the SCMF is well adapted to computing quadrupole matrix   
elements because of the large single-particle space that can      
be treated. The CISM requires severe truncation of the space,
and a systematic mapping procedure would determine the effective operators to be    
used in the smaller truncated shell-model space. 

Finally, such a mapping would be helpful to construct a global theory of
nuclear level densities. While the overall behavior of level densities can
be derived from the independent particle shell model, there are significant
interaction effects. The shell model Monte Carlo (SMMC) method has been
efficient for calculating level densities in large shell-model
spaces~\cite{na97,al99,al03}, but it requires as input a parameterized one-
and two-body shell-model Hamiltonian.  It should be mentioned that the
SMMC requires for its efficiency interactions that have a specific sign.\footnote{Interactions containing small terms having the bad sign can be
treated as well with the method of Ref.~\cite{al94}.} 
These conditions are satisfied for the Hamiltonian we will construct here.

Here we propose a mapping procedure and
apply it to $sd$ shell nuclei.  This region of
the table of nuclides is a paradigm for the CISM. For example, the USD 
Hamiltonian~\cite{wi84} defined in the $sd$ shell space 
reproduces very well binding energies,
excitation energies, and transition rates between states. 
For the actual SCMF that one might wish to map, we will use
one of the Skyrme family of energy functionals.  The corresponding SCMF equations
can be solved with computer
codes that are publicly available~\cite{st05,he05}; in particular we shall use
in this work the Brussels-Paris code~\cite{bo85,he05}, together with 
the SLy4 parameterization of the Skyrme energy functional~\cite{sly4}.
This functional is fit to a number of empirical properties of
nuclei and with a slight readjustment gives a 1.7 MeV rms
fit to masses of $\sim 600$ even-even nuclei~\cite{be05,ben05a}.

 The outline of this paper is as follows. In Section \ref{methodology} we discuss the main idea and procedure for constructing the SCMF to CISM map. Our present study focuses on quadrupolar correlations.  In Section \ref{USD} we test our mapping procedure using the mean-field approximation of the USD Hamiltonian as the SCMF theory to be mapped on a 
shell-model Hamiltonian. In Section \ref{mapping} we discuss the technical construction of the map, starting from the SCMF theory of the Skyrme family of energy functionals (in particular SLy4). In Section \ref{Si28} we carry out the mapping explicitly for the nucleus \Si,\, while in Section \ref{other-sd} we summarize similar results for all other deformed $N=Z$ $sd$ shell nuclei. In Section \ref{QQ-int} we use the mass quadrupole operator to construct an effective quadrupole-quadrupole interaction. Finally, in Section \ref{prospects} we discuss possible future developments of the theory introduced here.

\section{Methodology}\label{methodology}

Our goal is to choose a suitable truncated shell-model space and construct
within it a spherical shell-model Hamiltonian containing one- and two-body
terms.  We write this CISM Hamiltonian in the form\footnote{Throughout this work we will use a consistent notation with the following conventions. Operators
in Fock space (i.e., operators containing particle creation and annihilation operators)
are distinguished with a circumflex. Matrices representing single-particle
operators in a single-particle space are denoted in
boldface. The boldface is dropped for individual elements of those
matrices.}
\be\label{shell-H}
\hat H = \sum_{\alpha,j,m} \epsilon_{\alpha j} \hat a^\dagger_{\alpha j,m} 
\hat a_{\alpha j,m}
- {1\over 2}\sum_\kappa g_\kappa  \hat A_\kappa \cdot \hat A_\kappa \;.
\ee
Here  $\epsilon_{\alpha j}$ are single-particle energies of spherical 
single-particle orbitals.  The orbitals are characterized by the quantum numbers  $\alpha j m$, 
where $j,m$ is the total 
single-particle angular momentum and its projection, and by $\alpha$,
denoting the
remaining single-particle quantum numbers\footnote{E.g., $\alpha \equiv t_z, n, l$ with 
$t_z$ the isospin (proton or neutron), $n$ the radial quantum number and 
$l$ the orbital angular momentum}.
$\hat A_\kappa$ are one-body tensor operators and $g_\kappa$ are interaction
strengths. The index $\kappa$ incorporates
all the quantum numbers necessary to specify a spherical tensor operator of definite
rank $K_\kappa$.  In general
\be \label{multipole} 
\!\!\!\! \hat A_{\kappa M} = \sum_{\alpha j\alpha' j'}f^\kappa_{\alpha j\alpha' j'}\sum_{mm'}
 (j' m'  K_\kappa M | j m)\hat a^\dagger_{\alpha jm} \hat a_{\alpha' j'm'} \;,
\ee
and the scalar product\footnote{The tensor operator $\hat A_\kappa$ satisfies 
$ \hat A^\dagger_{\kappa M}=c(-)^{M}\hat A_{\kappa -M}$ with $c=\pm 1$ and is either 
hermitian ($c=1$) or anti-hermitian ($c=-1$). Therefore, the corresponding 
interaction term in (\ref{shell-H}) can also be written as 
$\hat A_\kappa \cdot \hat A_\kappa = c\sum_M \hat A^\dagger_{\kappa M} \hat 
A_{\kappa M}$.} in Eq.~(\ref{shell-H}) is $\hat A_\kappa \cdot \hat A_\kappa \equiv 
\sum_M (-)^M \hat A_{\kappa M} \hat A_{\kappa -M}\;.$ 

Eq.~(\ref{shell-H}) provides
the most natural form to which the SMMC method can be applied.
With enough terms in the $\kappa$ sum, it is
sufficiently general to treat any two-body interaction that is
rotationally invariant and time-reversal symmetric, 
The maximal number of terms $\kappa$ for any given rank
$K_\kappa$ of the multipole operators is the number of orbital
pairs whose angular momenta
$j,j'$ can be coupled to  $K_\kappa$ by angular momentum and
parity selection rules. We also note that
 $f^\kappa_{\alpha j, \alpha' j'}$ are proportional to the reduced matrix 
elements of $\hat A_\kappa$, i.e.,
$f^\kappa_{\alpha j, \alpha' j'}= 
(\alpha j||\hat A_\kappa||\alpha' j')/\sqrt{2 j +1}$.
In the development below we do not need to distinguish orbitals with
different quantum numbers $\alpha$, and we will drop that index.

Our starting point is the SCMF, which minimizes an energy functional for a
configuration of single-particle orbitals $\phi_k$ (i.e., a Slater
determinant) or the BCS generalization thereof.  Besides the ground-state
minimum, one can obtain constrained solutions by minimizing the energy
functional in the presence of an external field $\lambda \hat Q$, where
 $\hat Q$ is the mass quadrupole
operator defined below and
$\lambda$ is its strength.  The SCMF delivers a
set of orbitals $\phi_k^\lambda$ and corresponding single-particle energies
$\epsilon^\lambda_k$, as well as the total energy $E^\lambda$. Our task is
to use this information to construct a Hamiltonian of the form
(\ref{shell-H}).  The minimal requirement of such an effective Hamiltonian
is that it reproduces ground-state SCMF one-body observables sufficiently well within a mean-field approximation (in the truncated space).  We use this requirement as a guide to construct the effective Hamiltonian.

In the applications discussed in this work, we treat nuclei 
whose SCMF ground state is deformed.  This will simplify the 
mapping to some extent because we do not have to impose a constraining
field.  

The first step is the construction of a spherical
single-particle basis. Due to the rotational invariance of the
the energy functional, a spherically symmetric density matrix at a given iteration step  leads to a one-body Hamiltonian that is also spherically symmetric. For a
closed sub-shell nucleus (e.g., \Si) the density matrix in the next iteration step will remain spherically symmetric. Thus, for a closed sub-shell nucleus, if we start from a rotationally invariant single-particle density and iterate, we will
eventually converge to a spherical solution.  For an open shell nucleus,
we have to choose which $m$ orbitals of the valence partially
occupied $j$-shell are filled.  Such a choice will break the spherical
symmetry of the density and lead to a deformed solution. To overcome
this problem, one may use the uniform filling approximation
in which the last $n$ particles are distributed uniformly over
the corresponding $2j+1$ orbitals (i.e., we assume occupations
of $v_{j m}^2=n/(2j+1)$). As long as this approximation is used
in each HF iteration, we will converge to a spherical solution.
In practice, a pairing interaction or BCS-like occupations with
a fixed gap are used to guarantee the spherical symmetry of the
solution.

  We will discuss the construction of a spherical basis in more
detail in Section \ref{mapping}, but for now let us assume that we
have found the spherical solution with orbitals $\phi_{jm}$ as
well as the (deformed) ground-state solution with orbitals
$\phi_k$. The single-particle density matrix of the deformed
mean-field solution is given by $ \rhob_{\rm mf}$ with $
\rho_{{\rm mf},kk'}  = n_k \delta_{kk'}$, and $n_k$ is the 
occupation number of orbital $k$  
in the ground-state basis. In the HF approximation $n_k=1$ for
the lowest $N$ orbitals and $n_k=0$ for all other orbitals (here and in the following we use $N$ to denote the number of particles of a given type, i.e., either protons or neutrons).  We
transform single-particle operators to the spherical basis using
the transformation matrix $\bf U$ defined by taking the overlaps,
$U_{k,jm} \equiv \lb k | jm\rb$. Thus the single-particle
density matrix $\rhob$ in the spherical basis is given by
\be \label{rhojj'}
\rho_{jm,j'm'} = \sum_k U_{k,jm} n_k U_{k,j'm'} \;.
\ee
As long as the spherical basis $\phi_{j m}$ is
complete, $\bf U$ is unitary and (\ref{rhojj'}) is just a
different representation of the same ground-state density matrix
$ \rhob_{\rm mf}$. However, in constructing the mapping we choose
a truncated model space, and the matrix $\bf U$ (with $j m$ values
restricted to this space) is no longer unitary. This $\rhob$ is
then the density matrix in the truncated space, i.e., $ \bf P
\rhob_{\rm mf} P$, where $\bf P$ is the projector on the truncated
single-particle space.  We can calculate expectation values of
one-body observables in the {\em truncated} space by using this
truncated density $ \bf P \rhob_{\rm mf} P$.
For example, the expectation 
value of the mass quadrupole operator $\hat Q=\sum_{jm,j'm'}
q_{jm,j'm'} \hat a^\dagger_{jm} \hat a_{j'm'}$ in the truncated 
space may be computed as
\be \label{quad-moment}
\langle \hat Q \rangle = \tr ({\bf  q  P \rhob_{\rm mf} P})=   
 \sum_{jm,j'm'} q_{jm,j'm'}\rho_{jm,j'm'} \;,
\ee
where $q_{jm, j'm'} =\langle jm | \hat Q |j'm'\rangle$ are the matrix elements of
the mass quadrupole operator $\hat Q = \sum_i(2 z_i^2 - x_i^2 - y_i^2)$ in the single-particle
space.\footnote{In the following we will omit the projector $ \bf P$ in the
definition of the truncated $\rhob$ and it should be understood from the
context whether $\rhob$ denotes the complete or projected density matrix.}

Since the transformation $\bf U$ from the deformed basis into the truncated
spherical basis may not be unitary, it is important to choose the truncated
single-particle space such that the truncated ground-state density matrix
$\bf P \rhob_{\rm mf} \bf P$ satisfies some minimal properties required for
a density matrix that is the solution of the CISM Hamiltonian in a mean-field
approximation. One such requirement is that the number of particles be
correct; this is checked with the formula
\be
\label{N=tr-rho}
\tr \rhob = N\;.
\ee
This property is satisfied exactly in the full space with $N$
being the total number of particles, and the truncated
$\rhob$ should satisfy this condition approximately with $N$
being the number of valence particles.  Another
requirement is that the density matrix represent a single
configuration (i.e., a Slater determinant). This can be
expressed in the matrix equation
\be
\label{rho-sq=rho}
\rhob\rhob =  \rhob \;.
\ee 
 Equivalently, the requirement is that the eigenvalues of the density
matrix be zero or one (with exactly $N$ eigenvalues equal to one). 
 We will examine the eigenvalue spectrum of our
mapped density matrices to see how well this requirement is satisfied.

The mean-field Hamiltonian in the SCMF ground-state basis
is the single-particle matrix  $\bf h_{\rm mf}$ with elements
$ h_{{\rm mf},kk'} = \epsilon_k^0 \delta_{k,k'}$. Its matrix representation 
in the truncated spherical basis is given by 
\be\label{deformed-h}
{\bf h} = \bf U^\dagger {\bf h_{\rm mf} } U \;.
\ee
Let us expand $\bf h$ in multipoles, writing
${\bf h} = \sum_{K} {\bf  h}^{(K)}$, where  ${\bf h}^{(K)}$ is an
irreducible tensor of rank $K$. The reduced matrix elements of 
${\bf  h}^{(K)}$ are 
given by
\be
\label{h-K}
\!\! (j|| {\bf h}^{(K)}||j') = {2K+1 \over \sqrt{2j+1}} \sum_ {m,m'} (K M  j' m' | j m)
h_{jm,j'm'}
 \;.
\ee 

We now assume that the SCMF ground state is axially symmetry, so that only
the $M=0$ components of ${\bf h}^{(K)}$ are non-vanishing.  We shall
also restrict ourselves to even-even nuclei, which by time-reversal
invariance permits only even multipoles $K$.  The lowest multipole is the
monopole ${\bf h}^{(0)}$, from which we obtain the spherical single-particle
energies $\epsilon_j^{(0)}$, 
\be
\label{h-0}
  {h}^{(0)}_{jm,j'm'}= \epsilon^{(0)}_{j} \delta_{j j'}\delta_{m m'} \;.
\ee
These single-particle energies are given by $\epsilon^{(0)}_j=  \sum_m
h_{jm,jm}/(2j+1)$.

When the ground state is deformed, the $K=2$ part of the single-particle
Hamiltonian will be substantial and can be used to determine an effective
interaction in that quadrupolar channel. 
Defining the second-quantized one-body tensor operators (see Eq. (\ref{multipole}))
\be
 {\hat h}^{(K)}_M= \sum_{ j j'} {(j ||{\bf h}^{(K)}||j') \over \sqrt{2j+1}}\sum_{mm'}
 (j' m'  K  M | j m)\hat a^\dagger_{j m} \hat a_{j'm'} \;,
\ee
we consider the following effective CISM Hamiltonian
\be
\label{H-qq}
\hat H = \hat h^{(0)} - {1\over 2 } g \hat h^{(2)} \cdot \hat h^{(2)} \;.
\ee
The Hamiltonian (\ref{H-qq}) has the form (\ref{shell-H}) and
(\ref{multipole}) with a single multipole term $K_\kappa=2$ and ${\bf f} = {\bf h}^{(2)}$.

Assuming axial symmetry, the Hartree Hamiltonian of (\ref{H-qq}) is given by
${\bf h}_H= {\bf h}^{(0)} - g\langle \hat h^{(2)}_0\rangle {\bf h}^{(2)}_0$ (where 
$\lb \hat h^{(2)}_0\rb = \tr ({\rhob} {\bf h}^{(2)}_0)$). Choosing
$g$ to satisfy 
\be\label{Hartree-g}
{1\over g } = - \langle \hat h^{(2)}_0 \rangle\;, 
\ee
the Hartree Hamiltonian is ${\bf h}_H= {\bf h}^{(0)}+{\bf h}^{(2)}_0 $. 
If this procedure is carried out in the
complete space (i.e., no truncation), then the Hartree mean-field
Hamiltonian of (\ref{H-qq}) approximately coincides with the mean-field
Hamiltonian of the original SCMF theory, and their ground-state density
matrices coincide.  If $g$ is chosen to satisfy (\ref{Hartree-g}) in the
truncated space, then the Hartree Hamiltonian of (\ref{H-qq}) in the
truncated space is approximately the SCMF mean-field Hamiltonian projected
onto the truncated space. This will be strictly correct to the extent 
that $\langle \hat h^{(2)}_0 \rangle$ calculated with $\bf P  \rhob_{\rm mf}P$
(i.e., the SCMF density in the truncated space) coincides with 
$\langle \hat h^{(2)}_0 \rangle$ 
calculated with the self-consistent density matrix $\rhob$ of the 
Hartree problem in the truncated space.  

   Another important quantity is the total energy of the system.  The
absolute SCMF energies need not be reproduced by the mapped Hamiltonian,
but the energy difference between the mean-field
minimum $E_{\rm mf}$ and the spherical state $E_{\rm sph}$ should come out the same.
We
call this the deformation energy $E_{\rm def}$,
\be
E_{\rm def} = E_{\rm mf} - E_{\rm sph}.
\ee
A deformation energy can also be calculated by solving the CISM Hamiltonian (\ref{H-qq}) in the mean-field approximation. The above procedure to determine $g$ does not guarantee that the
deformation energy of the original SCMF theory is reproduced by the mapped
CISM Hamiltonian. We fine tune the value of $g$ so as to match the deformation energies of the SCMF theory and of the CISM Hamiltonian. As shown below, the deformation energy of the CISM Hamiltonian is very sensitive to $g$, and a relatively small change from its value determined by (\ref{Hartree-g}) will suffice to reproduce the correct deformation energy. On the other hand, the quadrupole moment and the ground density matrix (in the truncated space) are less sensitive to $g$ and are still expected to match approximately.

We have used the Hartree approximation to motivate our choice of the
effective interaction (\ref{H-qq}). However, in practice we include the
Fock term in the Hamiltonian to determine the density matrix
and deformation energy of the system. Using
$a \equiv (j m)$ to denote the spherical orbitals, the Hartree-Fock
Hamiltonian of (\ref{H-qq}) is given by
\begin{eqnarray}\label{HF}
h_{ac} = & h^{(0)}_{ac} - g \sum_M(-)^M h^{(2)}_{M;ac} \left(\sum_{bd} 
h^{(2)}_{-M;bd} \rho_{db}\right) \nonumber \\ 
& + g \sum_M (-)^M \sum_{bd} h^{(2)}_{M;ad}
h^{(2)}_{-M;bc} \rho_{db} \;,
\end{eqnarray}
where $h^{(2)}_{M;ac} \equiv \lb a |\hat h^{(2)}_M | c\rb$. If the mean-field
solution $\rho_{ac}$ in (\ref{HF}) possesses axial symmetry, only the $M=0$
component of $\hat h^{(2)}$ contributes to the direct term, while all values of $M$
can contribute to the sum in the exchange term.

    There are several sources for the imperfections in the mapping.
Besides the truncations in multipolarity and shell orbital space,
the underlying SCMF typically
has three-body interactions that are ignored in the mapped Hamiltonian.
Thus, it is important to check the mutual consistency of the mean-field
observables if the mapped Hamiltonian is to be used with confidence.  

\section{A warm-up exercise: the USD Hamiltonian}\label{USD}

\begin{figure}[bth!]

\centerline{\epsfig{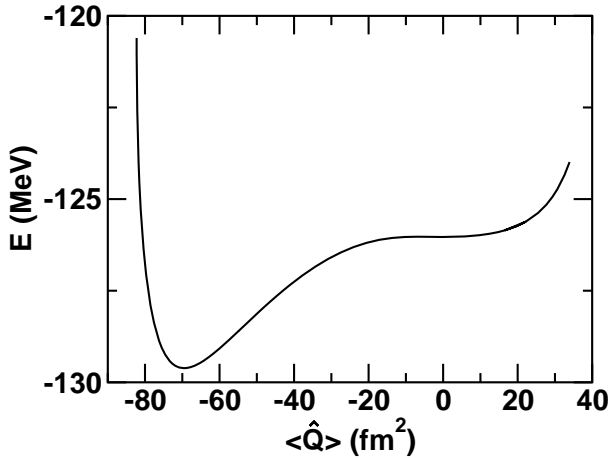}}
\caption{\label{si28_landscape}$^{28}$Si 
potential energy landscape obtained from the constrained Hartree-Fock
solution of the USD Hamiltonian. 
}
\end{figure}

As a warm-up exercise for the mapping onto the  
Hamiltonian Eq.~(\ref{H-qq}), let us consider the USD Hamiltonian~\cite{wi84} as the target and its mean-field approximation as the energy functional to
be mapped back onto an effective CISM Hamiltonian. As mentioned earlier, the
USD Hamiltonian gives a very good description of the properties of
$sd$-shell nuclei, including binding energies, spectra and transition
moments. We describe our procedure again but now taking the nucleus \Si~as
an example. First, we show the energy landscape for the solutions of
the constrained HF equation in Fig.~\ref{si28_landscape}. One
sees the ground-state minimum at an oblate deformation  
$\lb \hat Q \rb \approx -70$ fm$^2$ and a stationary point at zero deformation.  In fact we only need the properties of the HF solutions at these two points.

The spherical solution provides us with the spherical basis states for
expressing the deformed orbitals.  In the $sd$ shell-model space
this is trivial; the spherical single-particle HF states coincide with the original
single-particle states of the USD Hamiltonian and only their energies are
shifted.\footnote{The spherical state is always a stationary
solution of the mean-field equations (in the uniform filling
approximation), but it may not be lowest state satisfying the
constraint $\lb \hat Q\rb =0$. In that case, one has to find it in
some other way. In this section we have found the spherical solution by
starting from a spherical density and using the uniform filling
approximation at each iteration step.} In particular, a large
monopole contribution to the single-particle energies is
automatically subtracted when the spherical HF energies are
calculated.  The spherical state has a HF energy of $E_{\rm
sph}= -126.03$ MeV and zero quadrupole moment.

The deformed mean-field ground state is obtained by iterating the HF 
equations starting from a deformed oblate state and without a constraining
field.  The solution has an energy of
$E_{\rm mf}=-129.61$ MeV and a quadrupole moment of $\lb \hat Q
\rb=-20.3$ $b^2 = -69.6$ fm$^2$ (here $b$ is the
oscillator radius).\footnote{In the conversion from $b$ to fm we have used
$b^2=\hbar/m\omega_0$ with $\hbar \omega_0 = 45A^{-1/3} - 25 A^{-2/3}$.} The
deformation energy is the energy difference between the two solutions,
$E_{\rm def}= 129.61 - 126.03=3.58$ MeV.  The transformation matrix $\bf U$
is just the matrix describing the deformed mean-field orbitals in the shell
representation.  Since there is no orbital space truncation in the mapping
$sd\rightarrow sd$, the transformation matrix $\bf U$ 
is unitary and the consistency requirements on the ground-state density
matrix (\ref{N=tr-rho}) and (\ref{rho-sq=rho}) are automatically satisfied.

The next step is to express the 
single-particle Hamiltonian of the deformed solution in the shell representation,
Eq.~(\ref{deformed-h}), and carry out the multipole decomposition
keeping only the $K=0$ and $K=2$ terms.  This determines the
effective Hamiltonian Eq.~(\ref{H-qq}) except for an overall strength
parameter $g$.  

A simple estimate of $g$ can be obtained from the naive Hartree theory, 
$g=-1/\lb  \hat h^{(2)}_0 \rb = 0.0218$ MeV$/b^4$ (see Eq.~(\ref{Hartree-g})). 
We tune $g$ to match the deformation energy of the USD
Hamiltonian.  In practice we solve the HF equations for the
effective CISM Hamiltonian Eq.~(\ref{H-qq}) for a range of
values of $g$, making tables of the deformation energy
$E_{\rm def}(g)$ and the quadrupole moment $\lb \hat Q\rb $.  We then
determine a value for $g$ that fits the deformation energy of the USD
Hamiltonian $E_{\rm def}=3.58$ MeV. We find $g=0.0232$ MeV$/b^4$ and the
deformed state has a quadrupole moment of -18.2 $b^2$, about $10\%$ lower
than the actual minimum of the full USD Hamiltonian. 

\begin{table}[bth!]
\caption{\label{TableI} SCMF with the USD Hamiltonian and
the mapped Hamiltonian Eq.~(\ref{H-qq}) for $4N$ nuclei in the
$sd$ shell. Energies are in MeV, $\lb \hat Q\rb$ is in b$^2$ ($b$ being the oscillator radius) and $g$ is in MeV/b$^4$}

\begin{tabular} {c|cc|ccccc}
Nucleus & Interaction & $g$  & $E_{\rm sph}$  &  $E_{\rm mf}$  & $E_{\rm def}$ & 
$\langle \hat Q \rangle$ & $E_{\rm corr}$\\
\tableline
& USD  &  &-25.46 & -36.38 & 10.92 & 15.4 & 4.1\\
$^{20}$Ne & $\hat h^{(2)}$    &0.051 & -31.42 & -42.34 & fit & 15.0 & 3.5\\
\tableline
& USD  &  &-68.20 & -80.17 & 11.97 & 18.0 & 6.9\\
$^{24}$Mg & $\hat h^{(2)}$    &0.0261 & -97.04 & -108.98 & fit & 17.6 & 4.4\\
\tableline
& USD  &  &-126.03 & -129.61 & 3.58 & -20.3 & 6.3\\
$^{28}$Si & $\hat h^{(2)}$    &0.0232 & -182.11 & -185.72 & fit & -18.2 & 4.3\\
\tableline
& USD  &  &-222.75 & -226.56 & 3.82 & -13.5 & 4.0\\
$^{36}$Ar & $\hat h^{(2)}$    &0.062 & -372.75 & -376.55 & fit & -12.5 & 1.9\\
\end{tabular}
\end{table}

We have made similar calculations for the other deformed $N=Z$
even-even nuclei in the $sd$ shell. This excludes \S, which is
found to be spherical in the HF approximation.  The results are
summarized in Table I. We determine $g$ by matching the
deformation energy of the USD Hamiltonian, and in all cases we
find the quadrupole moment of the CISM Hamiltonian to be within
$\sim 3\% - 10\%$ of the quadrupole moment of the USD
Hamiltonian (where both moments are calculated in the HF
approximation).

\begin{figure}[bth!]
\centerline{\epsfig{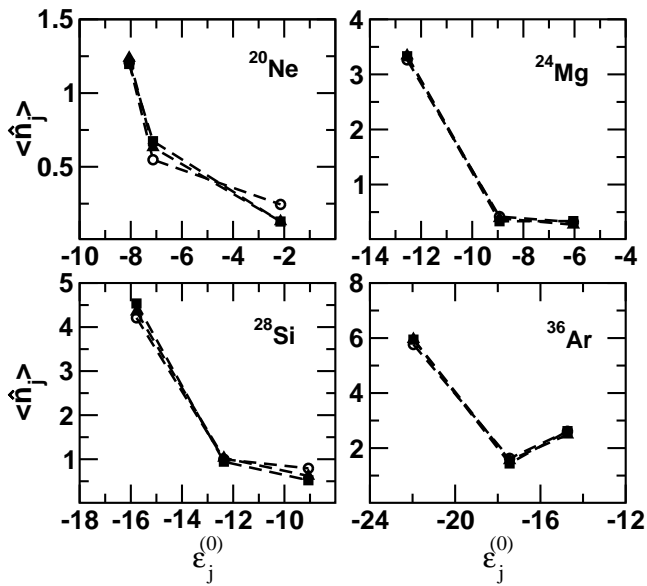}}
\caption{\label{sd-occupations}
Occupation numbers $\langle \hat n_j\rangle$ of the spherical orbitals in the deformed mean-field solution as a function of the spherical single-particle energy $\epsilon_j^{(0)}$ in four $sd$-shell nuclei (\Ne,\Mg, $^{28}$Si and \Ar). Results for the effective Hamiltonian (\ref{H-qq}) (solid squares) are compared with results for the USD interaction (open circles). Also shown are results for the effective interaction (\ref{QQ}) discussed in Section \ref{QQ-int} (solid triangles).
}
\end{figure}

In Fig.~\ref{sd-occupations}, we show the occupations $\lb 
\hat n_j\rb= \sum_m \lb \hat n_{jm} \rb$ of the spherical orbitals
versus the spherical single-particle energy $\epsilon_j^{(0)}$ in the deformed $4N$
$sd$-shell nuclei. The occupation numbers for the effective
Hamiltonians are comparable to the occupations for the USD
interaction.

\begin{figure}[h!]
\centerline{\epsfig{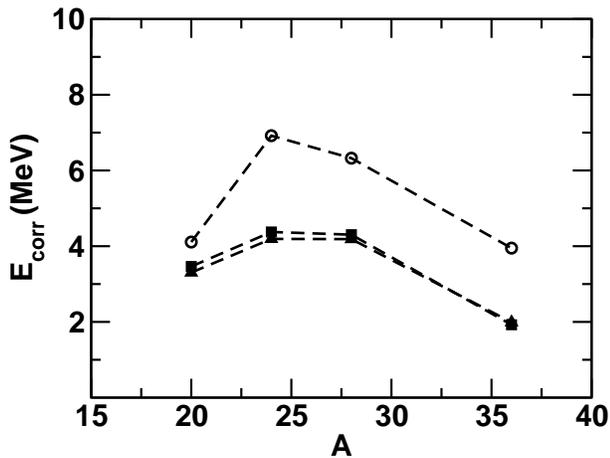}}
\caption{\label{sd_Ecorr}
Correlation energies $E_{\rm corr}$ versus mass number $A$ in deformed $N=Z$ $sd$ shell nuclei. The symbols are as in Fig.~\ref{sd-occupations}. Quadrupolar correlations are responsible for more than 50\% of the full correlation energies (with the exception of \nuc{36}{Ar}). 
}
\end{figure}

An important use of the effective CISM Hamiltonian will be to calculate the
correlation energy $E_{\rm corr}$, defined as the 
difference between the mean-field
energy $E_{\rm mf}$ and the CISM ground-state energy $E_{\rm gs}$,
\be
\label{eq:ecorr}
E_{\rm corr} = E_{\rm mf} - E_{\rm gs} \;.
\ee
The quadrupolar correlation energies extracted from the effective CISM
Hamiltonians are plotted versus mass number $A$ in 
Fig.~\ref{sd_Ecorr} (solid squares) and
compared with the full USD correlation energies (open circles). 
Note that the
quadrupole-quadrupole interaction is responsible for more than one half of
the total correlation energy in all cases but \nuc{36}{Ar}.  

\section{Mapping the SCMF} \label{mapping}

The SCMF of interest are based on local or nearly local energy functionals
and use large bases to describe the single-particle orbitals.  This
introduces technical problems in determining the spherical orbitals
as well as conceptual problems associated with the truncation of 
the space.  This section is devoted to these issues.  Some of the
details are specific to the {\tt ev8} SCMF code \cite{he05} that we
use for our numerical computations.

Since the effective CISM Hamiltonian $\hat H$ must be expressed in a spherical shell basis, we need to find the spherical
SCMF solution to construct the basis. We will denote the single-particle wave functions of this spherical solution by $\phi_{\alpha j m}$ and
truncate the space according to the needs of the CISM.
 Whether or not a solution is spherical can be easily
determined by the $2j+1$ degeneracy in the single-particle
energies. The $\lb \hat Q \rb=0$ HF ground state may or may not
be spherical.  Usually one includes a pairing field in the SCMF, which
stabilizes a spherical solution for $\lb \hat Q \rb =0$.   Our starting wave
functions are obtained from the archive \cite{archive}, which
were calculated including a delta-function pairing interaction
with a realistic strength.  We take the SCMF wave function
corresponding to $\lb \hat Q \rb = 0$ (which is spherical), and perform a few unconstrained iterations with the pairing interaction turned off. However, we include BCS-like orbital
occupation factors $v_k^2 =
(1- (\epsilon_k -\mu) / \sqrt{(\epsilon_k -\mu)^2 +
\Delta^2})/2$ with a fixed pairing gap $\Delta$ ($\mu$ is the chemical potential).  
This procedure is applied for several values of $\Delta$,
e.g., 2, 1 and 0.5 MeV. The resulting total spherical energies are found to have
an approximate linear dependence on $\Delta$ and we have used a
linear extrapolation to estimate the value at $\Delta = 0$. We take this extrapolated value to be the HF energy of the spherical solution.

 We use the code {\tt ev8}, which assumes that the SCMF solution
has time-reversal symmetry.  The orbitals then come in
degenerate time-reversed pairs, so only half of them are
considered, e.g., the positive
$z$-signature\footnote{The $z$-signature is the quantum number ($\pm 1$) that corresponds to the discrete symmetry transformation $e^{i \pi \left(\hat J_z -1/2 \right)}$.} orbitals~\cite{bo85}. For an axially symmetric solution, 
$J_z=m$ is a good quantum number and the positive $z$-signature 
orbitals have $m=1/2, -3/2, 5/2, -7/2,\ldots$. 
As a practical matter, there may not always be a clean
separation between different $m$ states (belonging to a given $j$-shell)
and even between orbitals of different $j$-shells, due to degeneracies and numerical
imprecision.   
In practice, we first determine $j$ by identifying, within the manifold of 
positive $z$-signature states, the $j + 1/2$ nearly degenerate 
states with a given parity. We then diagonalize the $(j+1/2)\times (j+1/2)$
matrix representing the axial quadrupole operator 
$ \hat Q^{(2)}_0= \hat Q$  to obtain states of good $|m|$. Using the Wigner-Eckart
theorem, we have
\be
\langle j m |\hat Q^{(2)}_0 | j m\rangle = c_j (j || \hat Q^{(2)} || j) [3 m^2 - j(j+1)]\;
\ee
with $c_j=2 [(2j-1) 2j(2j+1)(2j+3)]^{-1/2}$. Thus by arranging
the eigenvalues of $ \hat Q$ in ascending order, we identify
the orbitals with $|m|=j,j-1,\ldots,1/2$.\footnote{We note that
the reduced matrix element $(j|| \hat Q^{(2)} ||j)$ is negative so that
$\langle j j |\hat Q^{(2)}_0 | j j\rangle < 0$ is the lowest
eigenvalue. Indeed, in an orbital with good quantum numbers
$nlj$, $\langle n l j j |\hat Q^{(2)}_0 |n l j j\rangle = -{2j-1 \over
2j+1} \langle r^2 \rangle_{ nl} <0$. }

  Each good-$|m|$ orbital constructed above is determined up to an overall
phase. However, in the spherical shell model, the relative phases of the
states $|j m\rangle$ within a given $j$ manifold should be chosen such
that the states follow the standard sign convention, i.e., $J_\pm |j m\rangle
=\sqrt{j(j+1) - m(m\pm 1)} | j m\pm 1\rangle$. For the case with
time-reversal symmetry, the transformation matrix that diagonalizes
$ \hat Q^{(2)}_0$ is real. Thus each orbital $|j m\rangle$ is determined up to an
overall sign.

These signs can be determined by calculating the matrix elements of $ \hat Q^{(2)}_{
\pm 1}$ and requiring their sign to be consistent with the signs determined
by the Wigner-Eckart theorem
\be\label{Wigner-Eckart}
\langle j m |\hat Q^{(2)}_\mu | j m'\rangle = {(j m' 2 \mu | j m)\over \sqrt{2j+1}} (j || \hat Q^{(2)} ||j)
\ee
where $(j||\hat Q^{(2)}||j) <0$. Starting from the state $|m|=j$ and using the ladder
operators $\hat Q^{(2)}_{\pm 1}$, we can successively descend to states with
$|m|=j-1,\ldots,1/2$. At each step we determine the correct sign of the
state to be consistent with the sign of the corresponding Clebsch-Gordan coefficient in
(\ref{Wigner-Eckart}). We note that since only positive $z$-signature
orbitals are used, the sign of $m$ alternates in the above process and
therefore we need to calculate matrix elements of the type $\langle j m
|\hat Q^{(2)}_{\pm 1} | j \bar m' \rangle$ (where $\bar m$ is the time-reversed
state of $m$).

  In practice, we calculate the matrix elements $\langle j a |\hat Q^{(2)}_{\pm 1} | j \bar b\rangle$ in the original basis $|j a\rb$ obtained with {\tt ev8} (i.e., before diagonalizing $Q^{(2)}_0$) and then transform both $|j a\rangle$ and $|j \bar b\rangle$ to obtain the desired matrix elements
$\langle j m |\hat Q^{(2)}_{\pm 1} | j \bar m' \rangle$ (note that the same transformation matrix applies to both $|j a\rangle \to |j m\rangle$ and $|j \bar b\rangle \to | j \bar m \rangle$).

The {\tt ev8} code economizes on wave function storage by using
only the amplitudes associated with the first octant, $x,y,z
>0$. The symmetry properties of the wave functions under the $x=0$, $y=0$ and $z=0$ plane reflections should then be taken onto account in the evaluation of the quadrupole 
matrix elements; see the Appendix for details.  In the code and its output,
the orbital wave functions are represented on a lattice in coordinate
space. Once we have determined the standard spherical orbitals $\phi_{jm}$
and the deformed orbitals $\phi_k$, we can calculate the overlap matrix
$U_{k,jm}=\lb k |jm \rb$ by taking the scalar product of the respective
orbitals on the lattice.  The spherical-basis matrices of the (deformed)
ground-state density matrix $\rhob$ and the single-particle Hamiltonian ${\bf h}$
are calculated from (\ref{rhojj'}) and (\ref{deformed-h}), respectively.

\section{\Si~ in the SCMF}\label{Si28}

We now carry out the mapping for the SCMF of \Si~following the procedures
described in Sections \ref{methodology} and \ref{mapping}.  We use the SCMF
code {\tt ev8}, which takes as input a Skyrme parameterization of the
mean-field energy and a possible pairing interaction.  In this work we use
the SLy4 parameter set~\cite{sly4}.  The final energies in this work are
calculated in the HF approximation, with no pairing
contribution.\footnote{As discussed earlier, we may include a pairing
interaction when searching for the spherical solution but the energy is
extrapolated to the limit of no pairing.}

\begin{figure}[h!]
\centerline{\epsfig{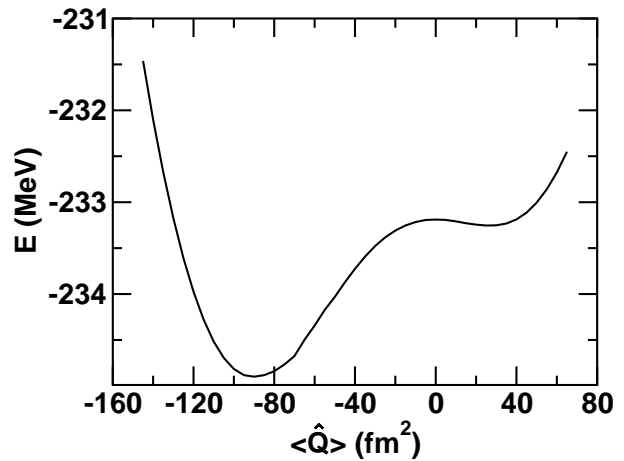}}
\caption{\label{si28-surface}
$^{28}$Si potential energy landscape obtained 
in the absence of pairing with the SLy4 interaction.
}
\end{figure}

The energy landscape of 
\Si~calculated using {\tt ev8} with a constraining mass quadrupole field
is shown in Fig.~\ref{si28-surface}.  Qualitatively, the landscape is
very similar to that found for the USD interaction in
Fig.~\ref{si28_landscape}.
The ground state at the minimum of the curve has an oblate deformation 
with a value $\lb \hat Q\rb = -89.2$ fm$^2$.  This is larger than the
value found for the USD interaction, which is to be expected because
the USD is a defined in a restricted space.  There are two other
stationary points in the landscape, namely the spherical saddle point and
and a shallow prolate minimum at $\lb \hat Q\rb \approx 30$ fm$^2$.  The deformation energy of the oblate solution $E_{\rm def}= E_{\rm mf} - E_{\rm sph}\approx 1.71$ MeV is lower than the deformation energy found for the USD interaction.

 A standard spherical basis $|j m \rb$ is constructed as described in Section \ref{mapping}. The reduced matrix elements of the mass quadrupole operator $\hat Q$ in the spherical basis are then extracted from
\be\label{rme-Q}
\! (j||\hat Q||j') = {5 \over \sqrt{2j+1}} \sum_m (2\; 0 \; j' m | j m) \lb j m
|\hat Q|j'm\rb
 \;.
\ee

\begin{figure}[h!]
\centerline{\epsfig{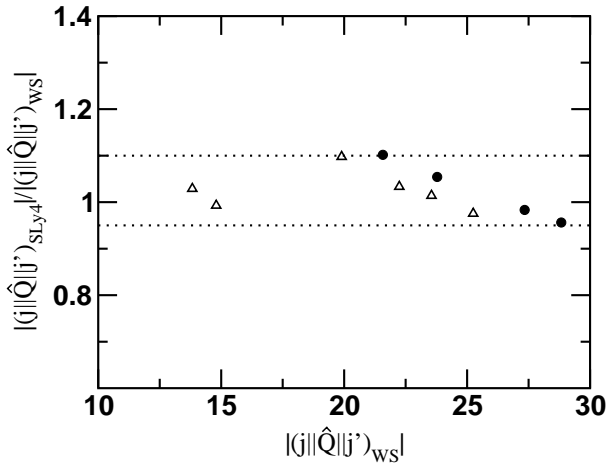}}
\caption{\label{si28_Q_ws} 
Ratio $|(j || \hat Q ||j')_{\rm SLy4}/(j || \hat Q ||j')_{\rm WS}|$ of the mass quadrupole  reduced matrix elements in $^{28}$Si versus 
$|(j ||\hat  Q ||j')_{\rm WS}|$. Solid circles: diagonal elements, open triangles: off-diagonal elements.
}
\end{figure}

 It is interesting to compare the reduced matrix elements (\ref{rme-Q})
to those of the simple spherical Woods-Saxon (WS) model, taking the parameters
of the Woods-Saxon plus spin-orbit potential from Bohr and 
Mottelson~\cite{bm69}. Fig.~\ref{si28_Q_ws} shows the ratio 
$|(j || \hat Q ||j')_{\rm SLy4}/(j || \hat Q ||j')_{\rm WS}|$ of the quadrupole
reduced matrix elements in the Sly4 SCMF and in the WS model versus
$|(j || \hat Q ||j')_{\rm WS}|$. 
We only show results for the valence $sd$ shell composed of the
$0d_{5/2}, 1s_{1/2}$ and $0d_{3/2}$ orbitals.  
The solid circles correspond to diagonal elements ($j=j'$) and open
triangles to off-diagonal elements. The values of this ratio vary
 in the range $\sim 0.95 - 1.1$ (dotted lines in Fig.~\ref{si28_Q_ws}). 
Thus, quadrupole renormalization effects in shell-model
calculations are more influenced by the truncation than by the
particular treatment of the spherical mean field.  Let us next examine those
truncation effects.

\subsection{Truncation effects and operator rescaling} 
\label{rescaling}

As mentioned earlier, the transformation from the deformed mean-field
orbitals to the spherical basis is not unitary owing to truncation
effects. The severity of the truncation can be assessed in several ways. The
first is the trace of the truncated single-particle density matrix, 
Eq.~(\ref{N=tr-rho}), which should equal the number of particles in the truncated
theory. With the truncation excluding orbitals beyond the
$N,Z = 20$ shell closures, we find the number of particles to be $27.75$ 
compared to $28$ for the full density matrix.  When truncated to the valence
$sd$ shell (i.e., without including the $0s_{1/2},0p_{3/2}$ and $0p_{1/2}$
core orbitals), the trace of the density matrix is 11.89 compared to $12$
valence particles in an $sd$ shell-model theory.  
A more severe test of the truncated single-particle density matrix 
is Eq.~(\ref{rho-sq=rho}), requiring that its eigenvalues be zero
or one.  Thus we diagonalize the truncated proton and neutron density matrices and 
examine the eigenvalues.  For the truncation to the 12+12 orbits of
the $sd$ shell, we find that all eigenvalues are within 1.5\% of the
required values.  
We conclude that the projected density $\bf P \rhob_{\rm mf} P$
satisfies to a good accuracy the HF condition for a Slater determinant.

\begin{figure}[t]
\centerline{\epsfig{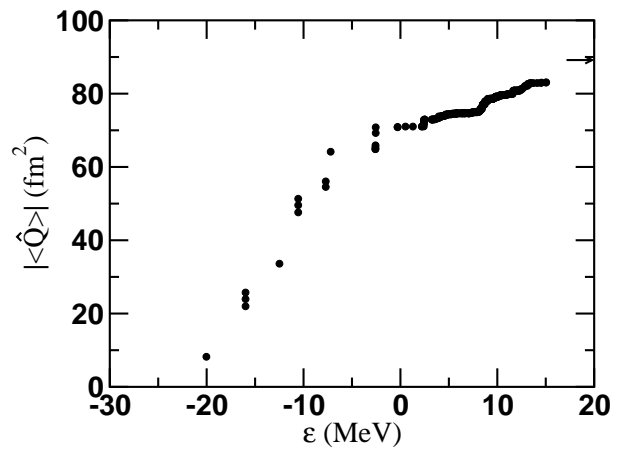}}
\caption{\label{si28_q_vs_ecutoff}
Quadrupole moment of the SCMF ground state of $^{28}$Si using the SLy4
Skyrme energy functional and different truncations in the spherical
basis.   The horizontal
axis is the maximum single-particle energy in the truncated space. Each solid circle  corresponds to a spherical single-particle state of protons or neutrons.
The untruncated moment of -89.2 fm$^2$ is shown by the arrow on the right.
}
\end{figure}

While the particle number is rather robust to truncation,
the same is not true of the matrix elements of the quadrupole 
operator.
To see how the effect of truncation evolves, we impose an upper cutoff energy
on the spherical orbitals and examine how the quadrupole moment changes with
this cutoff.  This is shown in Fig.~\ref{si28_q_vs_ecutoff},
describing $|\langle \hat Q\rangle |$ as a function of the maximal single-particle
energy used to define the spherical model space. With a 
cutoff of +15 MeV unbound orbital energy, the quadrupole moment is
reduced from its full value of -89 fm$^2$ by about 7\%.  Excluding all
positive energy orbitals,  the moment is -69 fm$^2$, a reduction of
$\sim 25\%$.  The truncation that excludes all orbitals above the $N,Z=20$ shell closure yields $\lb \hat Q \rb = -54.4$ fm$^2$. Finally, we find a slightly lower value of
$\lb \hat Q \rb = -52.4$ fm$^2$ when truncating to the valence $sd$ shell.  Thus, within a shell-model theory in the truncated valence $sd$ shell, the quadrupole operator should be rescaled by $89.4/52.4 \approx 1.71$.

\subsection{Mapped Hamiltonian}\label{mapped-H}

To obtain the mapped Hamiltonian, we first
calculate the transformation
matrix $\bf U$ between the deformed and spherical bases, using the output
orbital wave functions from {\tt ev8}. We then find the
deformed single-particle density matrix Eq.~(\ref{rhojj'}) and 
single-particle Hamiltonian Eq.~(\ref{deformed-h}) in the spherical basis. 
Next, we perform the multipole decomposition of the single-particle
Hamiltonian ${\bf h}= {\bf h}^{(0)} +  {\bf h}^{(2)} $ using (\ref{h-K}) and the Wigner-Eckart theorem (to convert the reduced matrix elements to matrix elements) to extract effective spherical shell orbital energies $\epsilon^{(0)}_j$ (see Eq.~(\ref{h-0})) and
and an effective quadrupole field ${\bf h}^{(2)}$. From this point on
the procedure is identical to what we did for the USD Hamiltonian.
We construct the effective CISM Hamiltonian (\ref{H-qq}) with a coupling constant $g$. A crude estimate for $g$ in the Hartree approximation is $g\approx - 1/\lb \hat h^2_0 \rb= 0.046$ MeV/fm$^4$. To determine $g$ more precisely we match the deformation energy of the effective CISM Hamiltonian (\ref{H-qq}) to its value $1.71$ MeV found in the SCMF theory (of SLy4). The deformation energy of the CISM Hamiltonian for a given value of $g$ is calculated from the spherical and deformed (oblate) solutions of (\ref{H-qq}) in the HF approximation. This procedure gives us $g=0.0594$ MeV/fm$^4$.

The mapped Hamiltonian can be tested by solving it in the HF approximation
and comparing with the SCMF. The deformation energy is the same by
construction. We obtain  a quadrupole moment of $-55$ fm$^2$, which 
is $\sim 5\%$ larger than
the SCMF quadrupole moment (projected on the $sd$ shell). We also find the
spherical occupations of the deformed HF ground state of the CISM
Hamiltonian to be rather close to the spherical occupations of the deformed
SCMF minimum (see Table IV below).

\subsection{Hartree approximation}

In the previous section, we solved the CISM Hamiltonian in
the HF approximation, i.e. including exchange.  In this section
we briefly examine the simpler Hartree approximation, because it 
allows some insight into the parametric sensitivities.  
Neglecting the exchange interaction, the dependence of the 
deformation energy $E_{\rm def}$
on the strength parameter $g$ can be conveniently determined in
two steps.  As a first step, the single-particle Hamiltonian
\be \label{hartree-h} 
{\bf h}_{\rm eff}= {\bf h}^{(0)} + \xi {\bf h}^{(2)}_0
\ee
is solved as function of a coupling parameter $\xi$, say in the
range $0<\xi<2$.  One needs to keep the
expectation values of ${\hat h}^{(0)}$ and ${\hat h}^{(2)}_0$ as a function of $\xi$, calculated with the ground-state density matrix $\rho_\xi$ of (\ref{hartree-h}).
For the second step, one determines the ground-state energy
$E(g)$ by finding the minimum in
\be
\label{Hartree}
E(g) = {\rm min}_\xi \left( \langle \hat h^{(0)}\rangle_\xi - {1 \over 2} g 
{\langle \hat h^{(2)}_0 \rangle}^2_\xi \right) \;.
\ee
The minimization in Eq.~(\ref{Hartree}) with respect to $\xi$ 
is equivalent to the self-consistent 
condition $g \langle \hat h^{(2)}_0 \rangle_\xi =\xi$.
The deformation energy is then given by
\be
E_{\rm def}(g) = E(0) - E(g)
\ee

\begin{figure}[ht]
\centerline{\epsfig{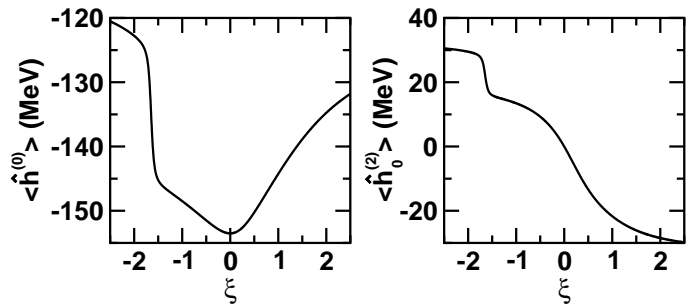}}
\caption{\label{e0_q_lambda}
Left: $\langle \hat h^{(0)} \rangle_\xi$ versus $\xi$  in $^{28}$Si. Right:  
$\langle \hat h^{(2)}_0\rangle_\xi$ versus $\xi$. The density matrix $\rhob_\xi$
corresponds to the single-particle (Hartree) Hamiltonian ${\bf h}_{\rm eff} =
{\bf h}^{(0)} +\xi {\bf h}^{(2)}_0$.}
\end{figure}

\begin{figure}[bth!]
\centerline{\epsfig{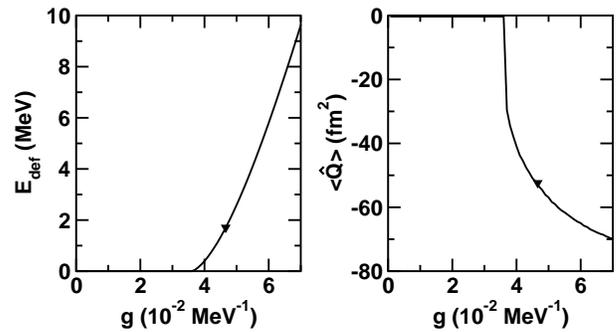}}
\caption{\label{si28_e0_q_g}
Left: deformation energy versus coupling parameter $g$ in $^{28}$Si when solving the effective Hamiltonian (\ref{H-qq}) in the Hartree approximation. Right: quadrupole moment 
 $\langle \hat Q\rangle$ versus $g$ in the Hartree approximation using the same effective Hamiltonian. The solid triangles describe the SCMF solution of the SLy4 energy functional (note that the triangle in the left panel is a fit).
}
\end{figure}

Fig.~\ref{e0_q_lambda} shows $\langle \hat h^{(0)}\rangle_\xi$ (left panel) and
$\langle \hat h^{(2)}_0 \rangle_\xi$ (right panel) versus $\xi$. The value of $\xi$ that
minimizes the Hartree energy is determined through the competition
between the two terms in Eq.~(\ref{Hartree}). 
Performing next the minimization to get $E(g)$, we find the deformation
energy curve shown in the left panel of Fig.~\ref{si28_e0_q_g}.  The panel
on the right in Fig.~\ref{si28_e0_q_g} shows the quadrupole moment of the
solution as a function of $g$. The discontinuity of $\langle \hat Q \rangle$ at
$g \approx 0.037$ MeV$^{-1}$ describes a first order spherical to deformed
shape transition.  Note that the deformation energy $E_{\rm def}(g)$ in the
deformed phase is a steep function of $g$; thus $g$ is well constrained by the
fitting procedure. The solid triangles in Fig.~\ref{si28_e0_q_g} indicate
the SLy4 SCMF point (in which the value of $g=0.0466$ MeV$^{-1}$ is fitted
to give the correct deformation energy).

\begin{figure}[h!]
\centerline{\epsfig{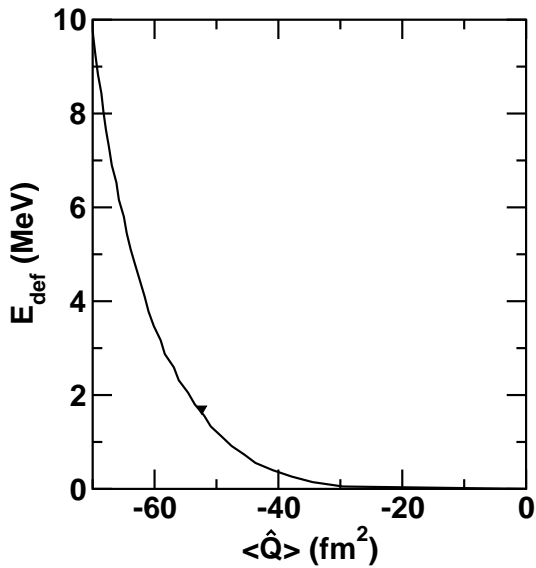}}
\caption{\label{si28_eq}
 Deformation energy versus quadrupole moment of $^{28}$Si using
  the Hartree approximation for the effective Hamiltonian (\ref{H-qq}) in
the $sd$-shell. The triangle shows the SCMF solution for the SLy4 energy
functional where the quadrupole moment is calculated in the
truncated space ($sd$ shell).
}
\end{figure}

The Hartree deformation energy $E_{\rm def}$ versus  $\langle
\hat Q \rangle$ in the truncated $sd$ space is plotted in Fig.~\ref{si28_eq}.
The solid triangle is the SLy4
SCMF point. It lies very close to the curve
extracted from the effective Hamiltonian, showing that the effective
theory closely reproduces the quadrupole moment in the truncated space. 

\subsection{Correlation energy}

With the mapped shell-model Hamiltonian in hand, we are now in a position
to compute the correlation energy as defined in Eq.~(\ref{eq:ecorr}).  The HF energy
 of (\ref{H-qq}) is found to be $E_{\rm mf} = -147.44 $ MeV. 
The ground-state energy $E_{\rm gs}$ is calculated using the shell-model code
{\tt oxbash}, putting in the orbital single-particle energies and interaction 
matrix elements from the mapped Hamiltonian (\ref{H-qq}).  We find $E_{\rm gs} = -153.93$ MeV.  Anticipating calculations in much larger shell-model spaces, we have also computed
$E_{\rm gs}$ with the SMMC.  Taking 
$\beta = 6 $ MeV$^{-1}$ and a time slice of $\Delta \beta=1/32$ MeV$^{-1}$ 
we find (using a stabilization method for the one-body propagation) $E_{\rm gs}= -153.87 \pm 0.07$  MeV, in agreement with {\tt oxbash}. 
For a time slice of $\Delta \beta=1/64$ MeV$^{-1}$ the SMMC result is $E_{\rm gs}=  -153.94 \pm 0.07$ MeV.  

The quadrupolar correlation energy we find for $^{28}$Si is then $E_{\rm
corr} \approx -4.55$ MeV.  This is quite comparable to the value we
obtained from a similar reduction of the USD Hamiltonian ($4.3$ MeV).  This gives
one some confidence that the quadrupole correlation energy has a
magnitude of about 4-5 MeV in \Si.  It is interesting to compare this 
with the quadrupole correlation energy obtained by a completely 
different approach, the generator coordinate method (GCM)~\cite{ben05}.  Using that
method in a global study of correlation energies~\cite{archive}, the authors
of Refs.~\cite{be05,ben05} obtained a correlation energy of 4.9 MeV  for \Si, 
which is somewhat larger.
However, these calculations included a pairing interaction, which we have not treated here. In particular, the $J=0$ wave functions used as the GCM basis are projected from deformed states that are constructed in the presence of a delta pairing force.      

\section{Other $\bf sd$ shell nuclei in the SCMF}\label{other-sd}

  We have repeated the mapping discussed in Section \ref{Si28} for the SCMF of all other deformed $N=Z$ nuclei in the $sd$ shell, i.e., for \Ne, \Mg~ and \Ar~ (\S~ is found to be spherical, just as for the USD interaction). The results are summarized in Tables II, III and IV (including results for \Si).

 Truncation effects on the deformed SCMF density matrix are summarized in
Table II. The eigenvalues and trace of the truncated density matrix 
$\PrhoP$  (projected on the valence $sd$ shell) are listed separately
for protons and neutrons (there are 6 positive $z$-signature eigenvalues for
each type of nucleon). We see that the HF condition $\rhob^2 =\rhob$
holds to a rather good accuracy for all four nuclei (all eigenvalues are
very close to either 1 or 0). The condition that the trace is given by the
number of valence particles is also satisfied quite well.

\begin{widetext}

\begin{table*}[bth!]
\caption{\label{TableII} Eigenvalues and trace of the truncated density matrix $\PrhoP$ in the SCMF theory of the SLy4 Skyrme energy functional. Shown are results for deformed $4N$ nuclei in the $sd$ shell (separately for protons and neutrons). }

\begin{tabular} {c|c|cccccc|c}

Nucleus &  & & &eigenvalues of $\PrhoP$ &  & & &$\tr(\PrhoP)$\\
\tableline
$^{20}$Ne & protons &  0.974 & 0.003 & 9 $\cdot 10^{-6}$ & 0 & -3 $\cdot 10^{-6}$ & -6$\cdot 10^{-6}$& 1.953\\
& neutrons & 0.975& 0.003 & 4$\cdot 10^{-6}$ & 0& -4 $\cdot 10^{-6}$& -6 $\cdot 10^{-6}$& 1.955 \\
\tableline
$^{24}$Mg & protons & 0.977 & 0.975 & 0.006 & 5$\cdot 10^{-6}$ & -2$\cdot 10^{-6}$& -8$\cdot 10^{-6}$ & 3.915\\
& neutrons & 0.977 & 0.975 & 0.006 & 7$\cdot 10^{-6}$ & -2$\cdot 10^{-6}$ & -6$\cdot 10^{-6}$&  3.916\\
\tableline
$^{28}$Si & protons & 0.991& 0.990& 0.987& 0.004& 7$\cdot 10^{-6}$& -6$\cdot 10^{-6}$&  5.944\\
& neutrons & 0.991& 0.991& 0.987& 0.004& 2 $\cdot 10^{-6}$& -8$\cdot 10^{-6}$ & 5.945 \\
\tableline
$^{36}$Ar & protons & 0.999 & 0.998 & 0.997 & 0.995&  0.993& 0.001 &  9.966 \\
& neutrons & 0.999& 0.998& 0.997&  0.995 & 0.994& 0.001 & 9.966   \\
\tableline
\end{tabular}
\end{table*}

\end{widetext}

\begin{table*}[t!]
\caption{\label{TableIII} SCMF with the SLy4 Skyrme energy functional and
the mapped Hamiltonian Eq.~(\ref{H-qq}) for deformed $4N$ $sd$ shell nuclei. All energies are in MeV, $\lb \hat Q \rb$ is in fm$^2$ and $g$ is in MeV$/$fm$^4$}

\begin{tabular} {c|cc|cccccc}
Nucleus & Interaction & $g$  & $E_{\rm sph}$  &  $E_{\rm mf}$  & $E_{\rm def}$ & 
$\langle \hat Q \rangle$ & $\langle \hat Q \rangle_{\rm mf}$ & $E_{\rm corr}$\\
\tableline
& SLy4  &  &-151.83 & -157.43 & 5.6 & 49.0 & 84.0& \\
$^{20}$Ne & $\hat h^{(2)}$    &0.077 & -30.17 & -35.78 & fit & 48.7 & & 2.39\\
\tableline
& SLy4  &  & -187.20 & -195.92 & 8.73 & 59.6 &111.9 & \\
$^{24}$Mg & $\hat h^{(2)}$    &0.0394 & -79.29 & -88.03 & fit & 60.6 & & 3.50\\
\tableline
& SLy4  &  &-233.19 & -234.90 & 1.71 & -52.4 & -89.2 & \\
$^{28}$Si & $\hat h^{(2)}$& 0.0594 &-147.67 &  -149.38  &    fit & -55.0 & & 4.55\\
\tableline
& SLy4  &  &-303.16 & -305.43 & 2.27 & -43.4 & -74.7 & \\
$^{36}$Ar & $\hat h^{(2)}$    &0.117 & -274.53 & -276.79 & fit & -42.7 & & 1.42\\
\end{tabular}
\end{table*}

\begin{table*}[bth!]
\caption{\label{TableIV} Spherical occupations $\lb \hat n_j\rb$ for deformed $4N$ nuclei in the $sd$ shell. The SCMF occupations (computed for the SLy4 Skyrme energy functional) are compared with the occupations for the mapped Hamiltonian Eq.~(\ref{H-qq}) both in the HF approximation and in shell-model calculations.}

\begin{tabular} {c|c|ccc|ccc}
Nucleus & Interaction &$p$ $0d_{5/2}$  &$p$  $1s_{1/2}$  & $p$ $0d_{3/2}$ & 
$n$ $0d_{5/2}$  & $n$ $1s_{1/2}$  & $n$ $0d_{3/2}$ \\
\tableline
          & SLy4       & 1.481  & 0.376 & 0.096 & 1.486 & 0.374 & 0.095\\
$^{20}$Ne & $\hat h^{(2)}$ (HF) & 1.565  & 0.372 & 0.064 & 1.569 & 0.366 & 0.065\\
          & $\hat h^{(2)}$ (CISM)& 1.531 &0.403& 0.066  &  1.539 & 0.396 & 0.065\\
\tableline
         & SLy4       & 3.290 & 0.411 & 0.215 & 3.293 & 0.408 & 0.215\\
$^{24}$Mg& $\hat h^{(2)}$ (HF) &  3.398  &   0.346   &  0.256  &   3.399  &  0.344 & 0.257\\
          & $\hat h^{(2)}$ (CISM)& 3.356 & 0.329 &  0.315 &   3.360  & 0.327  & 0.313\\
\tableline
        & SLy4       & 4.956  & 0.723 & 0.265 & 4.965 & 0.717 & 0.263\\
$^{28}$Si& $\hat h^{(2)}$ (HF) &  4.929 & 0.712 & 0.359 & 4.939 & 0.702 & 0.359 \\
          & $\hat h^{(2)}$ (CISM)& 5.011 & 0.633 & 0.356 & 5.010 & 0.633 & 0.357\\
\tableline
        & SLy4       & 5.919 & 1.751& 2.297 & 5.921 & 1.758 & 2.287\\
$^{36}$Ar& $\hat h^{(2)}$ (HF) & 5.979 & 1.694  & 2.327  & 5.980&  1.702 & 2.3185\\
          & $\hat h^{(2)}$ (CISM)& 5.971 & 1.662 & 2.367 & 5.971 & 1.668 & 2.361\\
\tableline
\end{tabular}
\end{table*}

Table III compares results for the SCMF with the SLy4 Skyrme force and for the HF of the mapped CISM Hamiltonian (\ref{H-qq}) with $g$ determined by matching the deformation energy of both theories. Shown are the spherical and mean-field total energies, deformation energy and the quadrupole moment $\lb \hat Q\rb$ in the truncated $sd$ shell-model space.  We observe that the quadrupole moment of the effective theory deviates by at most $5\%$ from its value in the SCMF (for \Si).

We also show in Table III the total SCMF quadrupole moment $\lb \hat Q \rb_{\rm mf}$  calculated in the deformed HF ground-state. Comparing $\lb \hat Q \rb_{\rm mf}$ with $\lb \hat Q\rb$ in the truncated $sd$ shell, we estimate that in the CISM theory the quadrupole operator should be rescaled by $\approx 1.71, 1.88, 1.70$ and $1.71$ for \Ne,\Mg,\Si~ and \Ar, respectively.

  The correlation energies (calculated from Eq.~(\ref{eq:ecorr}) with $E_{\rm gs}$ found using {\tt oxbash}) are shown in the last column of Table III.  The values we find are quite reasonable. They are below the values cited in~\cite{archive}, but we have not yet included pairing in our effective theory. 

Finally, in Table IV we compare the occupations of the spherical valence
orbitals of the SLy4 SCMF theory with similar occupations obtained with the
mapped Hamiltonian (\ref{H-qq}). The latter are calculated in the HF
approximation as well as in the shell-model theory from the expectation
values of $\hat n_j = \sum_m \hat a^\dagger_{jm} \hat a_{jm}$ in the correlated CISM
ground state.

\section{The $Q\cdot Q$ interaction}\label{QQ-int}

In nuclear physics there is a long history of modeling the effective interaction
as a quadrupole-quadrupole interaction of the form $\hat Q\cdot \hat Q$, 
including Elliott's SU(3) model
\cite{el58} and the pairing-plus-quadrupole model of Kisslinger
and Sorensen \cite{ki63}.  Since it is possible in some contexts
to exploit the properties of the quadrupole operator, it is 
of interest to see how well a quadrupole-quadrupole interaction performs in our context. 
 Thus, we apply the Hamiltonian
 \be \label{QQ}
 \hat H =\hat h^{(0)}  -{1\over 2} g \hat Q\cdot \hat Q \;.
 \ee
in the truncated shell-model space with the matrix elements of $\hat
h^0$ and the mass quadrupole operator $\hat Q$ determined by the appropriate mean-field theory. We follow the same procedure we used before to determine the coupling constant $g$.

\begin{figure}[h!]
\centerline{\epsfig{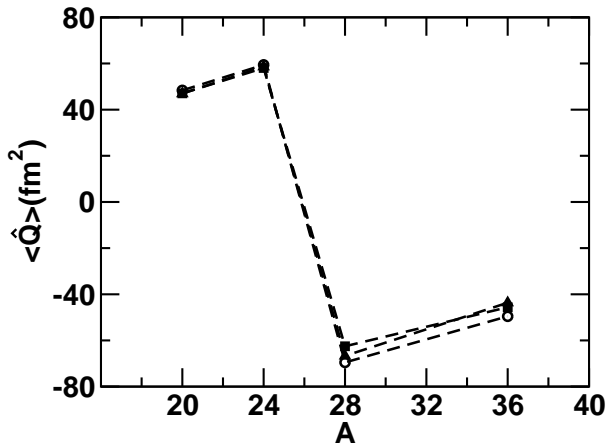}}
\caption{\label{sd_Q_A}
Quadrupole moment in the SCMF method versus mass number $A$ for $4N$
$sd$-shell nuclei. Results are shown for two effective CISM Hamiltonians:
the $\hat Q\cdot \hat Q$ interaction (\ref{QQ}) (solid triangles) and 
the $\hat h^{(2)}\cdot \hat h^{(2)}$ interaction (\ref{H-qq}) (solid squares). They are compared with
results for the USD Hamiltonian (open circles). 
}
\end{figure}

We have determined such an effective $\hat Q\cdot \hat Q$ interaction for the deformed $4N$ $sd$ shell nuclei \Ne,~\Mg,~\Si~and~\Ar~, using the HF approximation of the USD interaction as the SCMF theory.  We observe that the
quadrupole moments calculated with both effective interactions
 (\ref{H-qq}) and (\ref{QQ}) are quite
close to the USD results. This can be seen in Fig.~\ref{sd_Q_A}, where
$\langle \hat Q\rangle$ (in fm$^2$) is plotted versus mass number $A$. Results
for the $\hat Q\cdot \hat Q$ interaction (solid triangles) and the interaction with
$\hat h^{(2)}$ (solid squares) are compared with the USD results (open circles).
The effective Hamiltonians (\ref{H-qq}) and (\ref{QQ}) also yield similar occupations $\lb \hat n_j\rb$ (see Fig.~\ref{sd-occupations}) and correlations energies (see Fig.~\ref{sd_Ecorr}).

\section{Prospects and Conclusions}\label{prospects}

  We are encouraged by the results we found for constructing a
isoscalar quadrupolar interaction from the SCMF in the $sd$-shell, 
but much remains to be done to demonstrate that the method will
be useful in deriving a global theory of correlation energies.
We list some of the major tasks that should be addressed:

\begin{itemize}

\item
So far, we have carried out the mapping only for nuclei that have deformed
HF ground states.  If the lowest-energy mean-field solution is spherical, we
can use a constraining quadrupole field $\lambda \hat Q$ to deform the HF ground
state and use this deformed solution to construct the effective CISM
Hamiltonian. In particular, for a given strength $\lambda$ of the field, we
can compare the deformation energy of the constrained SCMF theory with the
deformation energy of the effective Hamiltonian in the constrained HF
approximation.  The Hartree-Fock Hamiltonian (\ref{HF}) of the effective
CISM Hamiltonian is now replaced by
\begin{eqnarray}\label{CHF}
&h^\lambda_{ac} =  h^{(0)}_{ac} - g \sum_M(-)^M h^{(2)}_{M;ac} \left(\sum_{bd} h^{(2)}_{-M;bd} \rho^\lambda_{db}\right) \nonumber
\\&  + g \sum_M (-)^M \sum_{bd} h^{(2)}_{M;ad} h^{(2)}_{-M;bc} \rho^\lambda_{db}+\lambda Q_{ac} \;,
\end{eqnarray}
where $\rho^\lambda_{ac}$ is the self-consistent density matrix in the
spherical basis (here $a=(j m)$) and $\hat Q$ is the axial quadrupole
operator. A practical problem is how large should we take $\lambda$ to be.
One possibility is to use the value of $\lambda$ that gives the deformed
$J=0$ minimum.\footnote{Nuclei with spherical HF minimum acquire deformation
after applying a $J=0$ projection \cite{be05}.}

\item  
The real challenge of a global theory is the heavy, open-shell nuclei,
and we should ask whether any severe problems will arise when we attempt
to map the SCMF onto the CISM Hamiltonian.  As a concrete example, consider
the nucleus \nuc{162}{Dy}.  It is strongly deformed, having
an intrinsic quadrupole moment around 18 bn.  The SCMF using the SLy4
interaction reproduces this value very well~\cite{ben05}. The 
deformation energy when pairing is not included is quite large (about $20$ MeV). 
The $J=0$ projection and quadrupole fluctuations (in the presence of a delta
pairing force) contribute an additional correlation energy of $\approx 3.4$
MeV~\cite{archive}. 

 To use an CISM Hamiltonian to calculate deformation and correlation energies, one
would need to truncate to a basis of the full shells, i.e., $N=82-126$ for
neutrons and $Z=50-82$ for protons plus possibly some intruder states. 
Thus, a first test would be to see how well the consistency conditions
Eqs.~(\ref{N=tr-rho}) and (\ref{rho-sq=rho}) are satisfied in such a truncated space. 
 
The next question that arises is the practicality of various implementations of the
CISM Hamiltonian for solving in such large spaces.  There are several approaches that
have promise, including the shell model Monte Carlo (SMMC) method~\cite{or94,al94,na97}, the
 Monte Carlo shell model~\cite{ot99} and direct matrix methods such as
those used by the Strasbourg group~\cite{ca03}.  One could also consider
more approximate theories of the correlation energy, such
as the RPA~\cite{st02}.

\item
  The mapping should be extended to include all other parts of the interaction that
we believe are well-determined in the SCMF, in particular
the interaction in the isovector monopole and quadrupole densities.  This
should be straightforward to carry out, requiring only that nuclei with
$N\ne Z$ be considered in a more general parameterization of the interaction.

\item
  The pairing interaction is very important in nuclear structure but is
notoriously difficult to specify in detail, i.e., its density dependence and
its spatial range (or equivalently the orbital space energy cutoff).  Most 
SCMF energy functionals include a pairing interaction, but only an 
overall strength can be reliably determined in phenomenological
theories.  Thus, absent an {\it ab initio} theory of the pairing 
interaction, we cannot expect to have a reliable decomposition of the pairing
fields in the mean-field Hamiltonian.  Still, it should be possible
to determine a strength parameter controlling the overall magnitude
of pairing effects.   Note that in the presence of pairing,
 the single-particle density matrices of the
SCMF theory have the BCS form.  Thus, to determine the parameters in the
mapped Hamiltonian, one would use the HFB or the HF-BCS approximations
for the self-consistent mean-field theory.

\item
  An unsatisfactory aspect of our mapping is the lack of unitary of the transformation from the deformed basis to the truncated spherical basis.  It would be useful to investigate mapping formalisms that yield
density matrices that satisfy exactly the criteria we have
discussed, i.e., Eqs.~(\ref{N=tr-rho}) and (\ref{rho-sq=rho}).

\end{itemize}

Despite this long list of tasks to be done, we believe
that the method discussed here can be developed to produce a well-justified global
theory of shell-model correlation energies, starting from a global SCMF theory (e.g., a Skyrme energy functional). Particularly encouraging is the finding that quadrupole correlations account for more than half the correlation energy in $sd$ shell nuclei.   

\section*{Acknowledgments}

We thank M.~Bender and P.H.~Heenen for instructions in using the SCMF code
{\tt ev8}. Y.A. would like to acknowledge the hospitality of the Institute of Nuclear Theory in Seattle where part of this work was completed. This work was supported by the U.S. Department of Energy under Grants
FG02-00ER41132 and DE-FG-02-91ER40608.

\section*{Appendix: wave functions and matrix elements in the
$z$-signature representation}

A deformed single-particle wave function in the density functional theory
 is a two-component complex
spinor $\psi =\left(\psi_+ \atop \psi_- \right)$, or alternatively a
four-component real spinor $\psi = \left( {{{\psi_1 \atop \psi_2} \atop
\psi_3} \atop \psi_4 } \right)$ with $\psi_1 ={\rm Re\,} \psi_+\;; \psi_2
={\rm Im\,} \psi_+\;; \psi_3 ={\rm Re\,} \psi_-$ and $\psi_4 ={\rm Im\,}
\psi_-$. The SCMF orbitals are chosen to have good parity $\pi$ and good
$z$-signature \cite{bo85}. Such states have well-defined symmetry
properties under each the three plane reflections $x\to -x;\; y \to -y;\; z
\to -z$ (see Table 1 in Ref. \cite{bo85}). Therefore it is sufficient to store their values in one octant (e.g., $x,y,z>0$).

The scalar product of two spinors $\psi$ and $\phi$ is a complex number with
${\rm Re \,} (\psi^\dagger \phi) = \sum_i \psi_i \phi_i$ and ${\rm Im \,}
(\psi^\dagger \phi )= \psi_1 \phi_2 - \psi_2 \phi_1 + \psi_3 \phi_4 - \psi_4
\phi_3$.

The find a standard basis of spherical single-particle orbitals $|j m\rb$,
 it is necessary to calculate matrix elements of the quadrupole operators $\hat Q^{(2)}_\mu$ (see Section \ref{mapping}).
 We denote the $j+1/2$ orbitals with good $j$ and positive $z$-signature by
$|j a\rangle$. The quadrupole matrix elements are obviously real and given by
\be\label{Q-matrix}
\langle j a |\hat Q^{(2)}_0 |j b\rangle = 8 \int_{x,y,z>0} dx\, dy\, dz\, (2 z^2 - x^2 -y^2)\, {\rm Re \,} (\psi^\dagger \phi)\;,
\ee
where $\psi$ and $\phi$ are the spinors describing $|j a\rangle$ and $|j b\rangle$, respectively. 

  To calculate $\langle j a |\hat Q^{(2)}_{\pm 1} | j \bar b\rangle$ we use the symmetry properties of the orbitals under plane reflections and find 
\begin{eqnarray}\label{xz-yz}
\langle j a | x z | j \bar b\rangle & = & 8 \int_{x,y,z>0} dx\, dy\, dz\, xz\, {\rm Re \,} (\psi^\dagger \bar \phi) \nonumber \\
\langle j a | y z |j  \bar b\rangle & = & 8 i \int_{x,y,z>0} dx\, dy\, dz\, yz \, {\rm Im \,} (\psi^\dagger \bar \phi) \;,
\end{eqnarray}
Since the time-reversed spinor of $\phi$ is given by
$\bar \phi = \left( \,\,\,\phi^*_- \atop -\phi^*_+ \right) = \left( {{{\,\,\,\phi_3 \atop -\phi_4} \atop -\phi_1} \atop \,\,\phi_2}
\right)$, we have
\begin{eqnarray}
 {\rm Re \,} (\psi^\dagger \bar \phi) & = & \psi_1 \phi_3 -\psi_3 \phi_1 +\psi_4 \phi_2 - \psi_2 \phi_4\nonumber \\
 {\rm Im \,} (\psi^\dagger \bar \phi) & = & \psi_3 \phi_2 -\psi_2 \phi_3 +\psi_4 \phi_1 - \psi_1 \phi_4 \;.
\end{eqnarray}
The matrix elements of $\hat Q^{(2)}_{\pm 1}$ are calculated from (\ref{xz-yz}) using $\hat Q^{(2)}_{\pm 1}= \mp \sqrt{6}\sum_i(x_i z_i\pm i y_i z_i)$ (they are all real).

\end{document}